\documentclass{appolb}
\usepackage{epsfig}

\begin{document}
\bibliographystyle{unsrt}
\title{Thermodynamic properties of nuclear matter \\ at finite temperature
\thanks{Presented by V. Som\`a at the XLVI Cracow School of 
Theoretical Physics}
}
\author{Vittorio Som\`a
\address{Institute of Nuclear Physics, PL-31-342 Cracow, Poland}
\and
Piotr Bo\.{z}ek
\address{Institute of Nuclear Physics, PL-31-342 Cracow, Poland\\
and\\
Institute of Physics, Rzeszow University, PL-35-959 Rzeszow, Poland}
}
\maketitle
\begin{abstract}
Thermodynamic quantities in symmetric nuclear matter are calculated
in a self-consistent $T-$matrix approximation at zero and
finite temperatures. The internal energy is calculated from the 
Galitskii-Koltun's sum rule and from the summation of the diagrams for the 
interaction energy. 
The pressure and the entropy at finite temperature are obtained
from the generating functional form of the thermodynamic potential.
\end{abstract}

\section{Introduction}
Nuclear matter is studied as a many-body system of interacting fermions.
Its equation of state is of particular interest because of its applications
to heavy-ion collisions, supernovae collapses and neutron stars structure.
The thermodynamic properties of such systems can be investigated starting
from the nucleon-nucleon potential; strong interactions between nucleons
induce short range correlations in the dense nuclear matter. 

Our approach is based on finite-temperature Green's functions. 
We adopt the $T-$matrix approximation \cite{kadBaym}, which consists in 
expressing the self-energy as a sum of ladder diagrams.
Such a self-consistent scheme
belongs to a class of approximations derivable from a suitably chosen
generating functional \cite{kadBaym,baym}, which automatically fulfill
thermodynamic relations, including the 
Hugenholz-Van Hove  and Luttinger identities \cite{hvh,luttinger2}.

For what concerns the binding energy at zero temperature the 
thermodynamically consistent $T-$matrix approximation 
\cite{Bozek:2002tz,Dewulf:2003nj}
yields results similar to other approaches, such as extensive variational and 
Brueckner-Hartree-Fock calculations \cite{vcs2,Baldo:2001mw}.
It is hazardous to compute the pressure and the entropy simply by integrating
thermodynamic relations because of uncontrollable numerical errors that
propagate at small temperatures. We present instead a method which allows to
reliably calculate the pressure (and consequently the entropy) at sufficiently 
high temperatures from the diagrammatic expansion of the thermodynamic 
potential \cite{soma-bozek}.

The starting point of these studies is a free
N-N potential, a parameterization of the interaction derived from
nucleon-nucleon scattering experiments. In the present
paper we restrict ourselves to the two-body CD-Bonn potential and will
consider three-body interactions in a future work.

\section{$T-$matrix approximation}
In the in-medium $T-$matrix approach the two-particle
propagator takes the following form (for simplicity we skip spin and isospin 
indices, as well as energy and momentum dependence)
\begin{equation}
\label{app}
\mathcal{G}_{2}=G_2^{nc}+G_2^{nc} \: T \: G_2^{nc}
+ \: exchange \: terms \: \: .
\end{equation}
$G_2^{nc}$ is the  \textit{non correlated} two-particle Green's function,
just constructed as a product of two dressed single-particle propagators.
We remark that
the use of the dressed propagator implies the presence of a nontrivial
dispersive self-energy which leads to a broad spectral function
\cite{Bozek:2002em}. 
The in-medium two-particle scattering matrix $T$, which appears in the above
expression, is defined as 
\begin{equation}
T=V+V\:G_2^{nc} \: T
\end{equation}
where $V$ is the interaction potential. All the
ingredients are then calculated iteratively: the scattering matrix, the
single-particle self-energy, which is expressed in terms of
the scattering matrix, and the single-particle propagator, by means of the
Dyson equation.

The  $T-$matrix approximation scheme can be as well derived from a 
generating functional $\Phi[G,V]$ which is a set of two-particle irreducible
diagrams~\cite{baym} obtained by closing    two-particle    ladder
diagrams  with a combinatorial factor $1/2n$, where $n$ is the number of
interaction lines in the diagram. 
\begin{figure}[h]
\begin{center}
\includegraphics[width=10cm]{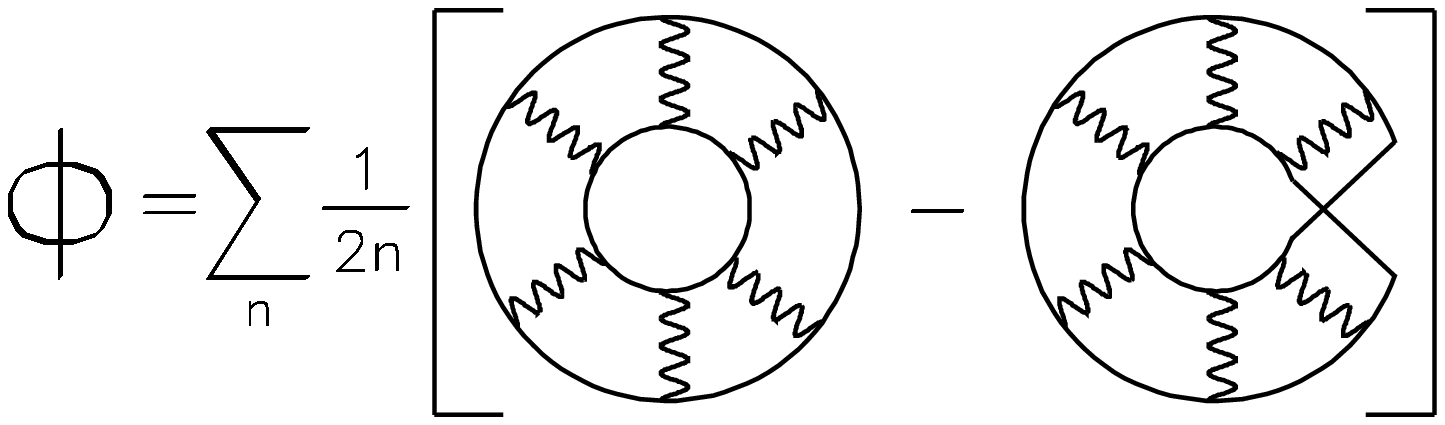}
\caption{Two-particle irreducible functional for the $T-$matrix approximation.
The sum runs over the number of interaction lines $n$ in the two-particle 
ladders.}
\label{fig:tphi}
\end{center}
\end{figure}
 The functional $\Phi$ depends on the dressed
propagators (lines in Fig. \ref{fig:tphi})
 and on the two-particle potential (wavy lines 
  in Fig. \ref{fig:tphi}).
The self-energy is then expressed as a functional derivative 
$\displaystyle \Sigma=\frac{\delta\Phi}{\delta G}$.

\section{Properties of the correlated system}
\subsection{Internal energy}
The (total) internal  energy per particle can be calculated as the expectation
value of the Hamiltonian of the system $H = H_{kin} + H_{pot}$
\begin{equation}
\label{eq:be}
\frac{E}{N} = \frac{1}{\rho} 
\left [
\frac{\langle H_{kin} \rangle}{\mathcal{V}} + 
\frac{\langle H_{pot} \rangle}{\mathcal{V}} 
\right ]\: .
\end{equation}
Here $\rho$ is the density and $\mathcal{V}$ the volume of the system.
When we evaluate the right hand side of (\ref{eq:be}) in momentum space
we see that the kinetic part is 
\begin{equation}
\label{eq:kin}
\langle H_{kin} \rangle = \mathcal{V} \int \frac{d^3{p}}{(2\pi)^3} 
\frac{d \omega}{2\pi}
\frac{\mathbf{p}^2}{2m} A(\mathbf{p},\omega) f(\omega) \: ,
\end{equation}
where $A(\mathbf{p},\omega)$ is the spectral function and
$f(\omega)$ the Fermi-Dirac distribution.
For the potential term, using the approximation 
(\ref{app}) for the two-particle Green's function, we get 
\begin{eqnarray}
\label{eq:final}
& \displaystyle
\langle H_{pot} \rangle = \frac{\mathcal{V}}{2} 
\int \frac{d^3{P}}{(2\pi)^3} \frac{d^3{k}}{(2\pi)^3} 
\frac{d\Omega}{2\pi}  \, b(\Omega) \, 
\nonumber \\
& \displaystyle \times
\mbox{Im} \left\{ \left( \langle \mathbf{k}|T^R(\mathbf{P},\Omega)
|\mathbf{k}\rangle-
 \langle \mathbf{k}|T^R(\mathbf{P},\Omega) |\mathbf{-k}\rangle\right)\,
{G_2^{nc\ R}}(\mathbf{P},\mathbf{k},\Omega)\right\} \: ,
\end{eqnarray}
where $b(\Omega)$ is the Bose-Einstein distribution. The dependence on the 
total energy $\Omega$ and momentum $\mathbf{P}$ of the pair and on the 
exchanged momenta $\mathbf{k}$ is now made explicit.

It is also possible to determine the energy in an alternative and simpler
way. The Galitskii-Koltun's sum rule \cite{galitskii,martinschwinger,koltun}
\begin{equation}
\label{gksumrule}
\frac{E}{N} = \frac{1}{\rho}
\int \frac{d^3{p}}{(2 \pi)^3} \frac{d \omega}{2 \pi}
\left [ \frac{\mathbf{p}^2}{2m} + \omega \right ] 
A(\mathbf{p},\omega) f(\omega )
\end{equation}
for conserving approximations is 
equivalent to the direct calculation (\ref{eq:be}).
However this is valid when only two-body interactions are considered.
In the presence of many-body forces the sum rule cannot be employed and the
internal energy of the correlated system has to be computed from the
expectation value of the Hamiltonian (\ref{eq:be}). 

\begin{figure}[h]
\begin{center}
\includegraphics[width=13cm]{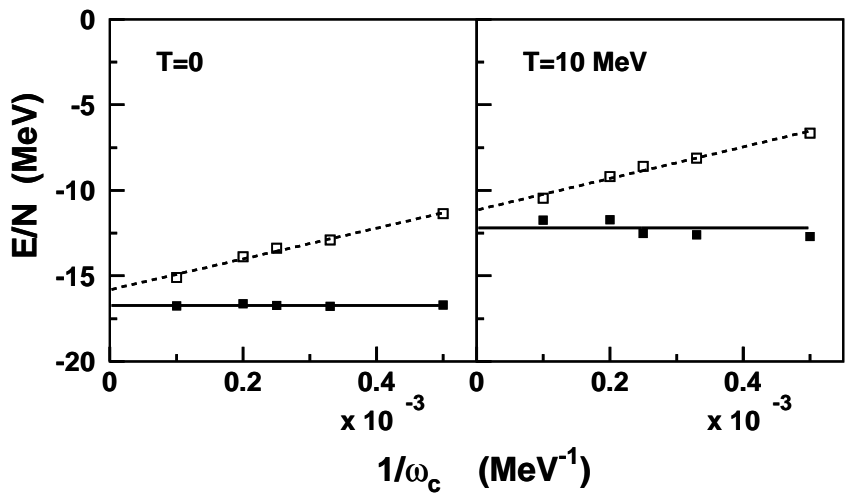}
\caption{Internal energy per particle as a function of the cutoff at
$T=0$ and $T=10\,\mbox{MeV}$. The dashed lines represent the energy
calculated from the sum rule (\ref{gksumrule}), the solid lines the energy
from the summation of diagrams.}
\label{fig:doublecutoff}
\end{center}
\end{figure}

We calculate the internal energy per particle with the use of the two methods:
from the Galitskii-Koltun's sum rule (\ref{gksumrule}) and
from diagram summation, i.e. from eq. (\ref{eq:be}) together 
with (\ref{eq:kin}) and (\ref{eq:final}).
We restrict ourselves to the empirical saturation density 
($\rho=0.16$ fm$^{-3}$) and consider different temperatures up to 
$20\,\mbox{MeV}$.
The calculation are performed using a numerical procedure where the energy 
range  is limited to an interval $[-\omega_c,\omega_c]$. We have tested
several values of the energy scale $\omega_c$ between
$2\,\mbox{GeV}$ and $10\,\mbox{GeV}$, and we found 
that the Galitskii-Koltun's sum rule expression for the energy shows some
dependence on this value. On the other hand, the results obtained from the 
direct estimation of the interaction and kinetic energies are stable and
independent on the energy range taken 
(up to an inaccuracy due to numerical discretization 
of about $0.5$ MeV). 
The cutoff dependence is shown in Fig. \ref{fig:doublecutoff} for two
different temperatures.
The result from the diagram summation can be compared to 
Galitskii-Koltun's sum rule result extrapolated to infinite energy range
(Table \ref{tab:pressure}).
This extrapolated value will be used in the following.

\subsection{Pressure}
The pressure is related to the thermodynamical potential through
\begin{equation}
\label{th-pre}
\Omega(T,\mu,V)=-P \: \mathcal{V} \: .
\end{equation}
It can be shown that
\begin{equation} 
\label{om}
\Omega=-\mbox{Tr}\{\ln[G^{-1}]\} -\mbox{Tr}\{\Sigma G\} + \Phi \: .
\end{equation}
The first two terms involve an integration over energy and momenta of
single-particle propagators and self-energies. The calculation of the
functional $\Phi$ requires instead a summation of a set of diagrams,
evaluated in the following way.
One notices that within the $T-$matrix approximation 
the diagrammatic expansions for the interaction energy and the 
functional $\Phi$ differ by the factor $1/n$ where $n$ is 
the number of interaction lines in the diagram 
. 
Hence the functional $\Phi$ can be 
obtained from the formula for the interaction energy
\begin{equation}
\label{eq:ourphi}
\Phi=\int_0^1
\frac{d\lambda}{\lambda} <H_{pot}(\lambda V, G_{\lambda=1})> \ \ .
\end{equation}
In the above formula  the interaction potential is multiplied by the 
factor $\lambda$ but the
propagator $G$ is the dressed nucleon propagator corresponding to the
system with the full strength of interactions $(\lambda=1)$. 
\begin{table}[htbp]
\begin{center} 
\begin{tabular}{|c||c|c|c|c|} 
\hline $\; T   \;$  & $\; E_{GK}/N \;$ & $\;  E_{diag}/N \;$ &
$\; P_{tot}/ \rho  \;$ &  $\; P_{BF}/\rho \; $
\\ \hline 
\hline $0$ & $-15.80$ & $-16.63$ &  $-7.69$ & $-5.58$ 
\\ 
\hline $2$ & $-15.15$ & $-16.29$ & $-5.86$ & $-5.4$
\\ 
\hline $5$ & $-14.40$ & $-15.24$ &  $-5.48$ & $-5.2$
\\ 
\hline $10$ & $-11.15$ & $-11.72$ & $-2.66$ & $-3.35$
\\
\hline $20$ & $-1.29$ & $-1.21$ & $\:\:6.49$ & $4.74$
\\
\hline
\end{tabular} 
\end{center} 
\caption{Results (at $\rho = \rho_0$) of the internal energy and the  pressure 
at zero and finite temperature
(all values in MeVs). The second and third colum represent the internal energy
per particle obtained from the Galitskii-Koltun's sum rule (\ref{gksumrule})
and from the diagram summation (\ref{eq:be}).
The other columns contain our estimate for the pressure 
(from eq. (\ref{th-pre})) and
the results of Baldo and Ferreira \cite{Baldo:Ferreira}.}
\label{tab:pressure} 
\end{table}

The pressure in hot nuclear matter has been obtained using
two-body Argonne $v_{14}$ interaction in the Bloch-De Dominicis approach 
\cite{Baldo:Ferreira}. The results are qualitatively similar, with a negative
value of the pressure at $T=0$ and $\rho\simeq 0.16$ fm$^{-3}$. This shows
the need to include three-body forces for a reliable description of the
thermodynamics of the nuclear matter.

\subsection{Entropy}
\begin{figure}[h]
\begin{center}
\includegraphics[width=8cm]{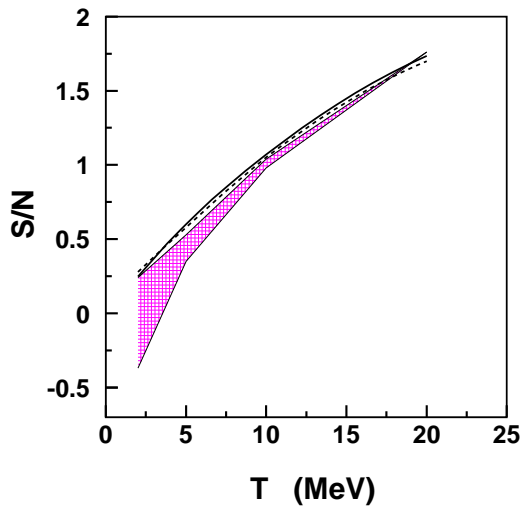}
\caption{Entropy per baryon 
as a function of the temperature. The hatched band denotes our 
estimate of the entropy. The solid line denotes the result 
for the free Fermi gas with the in-medium effective mass and the dashed line
represents the result of \cite{barcelona}  (Eq. \ref{eq:ssimp}).}
\label{fig:entropy}
\end{center}
\end{figure}

We compute the entropy through the thermodynamic relation
\begin{equation}
\frac{S}{N} = \frac{1}{T} 
\left [
\frac{E}{N} + \frac{P_{tot}}{\rho} - \mu
\right ] \: .
\end{equation}
The results are plotted in Fig. \ref{fig:entropy}.
The error in the calculation  of $TS$ at each temperature can be estimated by
comparing the results obtained with the two expressions for 
the internal energy 
(\ref{eq:be}) and (\ref{gksumrule}). The difference is of the order of
$1$ MeV, also at zero temperature we find $|TS|\simeq 1\,\mbox{MeV}\neq 0$.
The entropy can be estimated reliably only for $T\ge 5$~MeV, with the
uncertainty shown as the hatched band in the figure.

We compare these results with other methods of calculating the entropy:
\begin{enumerate}
\item The dynamical quasiparticle formula \cite{pethickentropy}
\begin{eqnarray}
\label{eq:ssimp}
& \displaystyle  \frac{S_{DQ}}{N} =\frac{1}{\rho}
 \int \frac{d^3p}{(2 \pi)^3} \frac{d \omega}{2 \pi}
\sigma(\omega) \left[A({\bf
 p},\omega)\left(1-\frac{\partial\mbox{Re}\Sigma^{R}
(\mathbf{p},\omega)}
{\partial\omega}\right)\right.
\nonumber \\
& + \displaystyle \left.
\frac{\partial\mbox{Re}G^{R}
(\mathbf{p},\omega)}
{\partial\omega}\Gamma(\mathbf{p},\omega)\right] \ \ ,
\end{eqnarray}
where
\begin{equation}
\sigma(\omega) = - f(\omega) \ln [f(\omega)] 
- [1-f(\omega)] \ln [1-f(\omega)] \: .
\end{equation}
It has been shown that this one-body formula gives results close to the 
complete expression for the entropy \cite{barcelona}.
\item The entropy calculated as for a free Fermi gas but using the effective
mass $m^*$ (determined at each temperature by
${(\partial {\omega}_p / \partial p^2)}_{p=p_F}=1/2 m^\star$)
instead of the rest mass $m$
\begin{equation}
\label{eq:ffge}
\frac{S_{{free}^\star}}{N} =\frac{1}{\rho} \int \frac{d^3p}{(2\pi)^3}\
\sigma\left(\frac{p^2}{2m^\star}\right)   \ \ .
\end{equation}
\end{enumerate}
The entropy estimated by means of (\ref{eq:ssimp}) and (\ref{eq:ffge}) 
is shown in Fig. \ref{fig:entropy}
with the dashed and solid line respectively.
Remarkably, we find that the expression for 
the entropy of the free Fermi gas is very similar to the result of the full
calculation for the interacting system, if the
change of  the effective mass in the system is taken into account.
The last observation  simplifies  significantly 
 the modeling of the evolution of protoneutron stars 
\cite{ Prakash:1996xs}, since relations between the entropy
per baryon and the temperature derived for a fermion gas 
can be used in hot nuclear matter. As observed in \cite{barcelona} the 
quasiparticle expression (\ref{eq:ssimp}) for the entropy 
\cite{pethickentropy} follows closely the full result.

\section{Summary}
We investigate the properties of correlated nuclear matter up to $T=20$~MeV. 
We calculate the internal energy with different methods, through the 
Galitskii-Koltun's sum rule and by summing the diagrams which contribute
to the expectation value of the interaction energy. The two methods give
similar results at all temperatures up to a difference of about $1$ MeV 
which can be attributed to numerical inaccuracies.
The pressure is derived from the thermodynamic potential which is given by
the generating functional $\Phi$. The
diagrams contributing to $\Phi$ are calculated with an integration 
of the interaction energy over an artificial parameter $\lambda$ which
multiplies the interaction lines, while keeping the propagators dressed as in
the fully correlated system. 
We finally estimate the entropy and compare it to two expressions derived
in the context of a quasiparticle gas, whose results turn out to be close 
to the full calculation.

Only a study which embodies
three-nucleon interactions can be realistically compared to
experimental data on the nuclear equation of state. Here all calculations
are carried out at the empirical saturation density, but
the present scheme can be applied to other densities and it
is suitable for the inclusion of three-body forces
or the study of asymmetric nuclear matter, which will be addressed in a
future work.

\bibliography{pibib}

\end{document}